\documentclass[a4paper,fleqn,usenatbib,useAMS]{mnras}
\usepackage{graphics}
\usepackage{graphicx}
\usepackage[pagewise]{lineno}
\usepackage{amsmath}    
\usepackage{amssymb}    
\usepackage{multicol}        
\usepackage{bm}     
\usepackage{pdflscape}  
\usepackage{xcolor}
\usepackage{ae,aecompl}
\usepackage{subfigure}
\usepackage{scalerel}
\usepackage[english]{babel}
\usepackage{times}
\usepackage{physics}
\usepackage{etoolbox}
\usepackage{multirow, booktabs}
\usepackage{siunitx}
\usepackage{paralist}
\usepackage{makecell}
\usepackage{tabularx}
\usepackage[inline]{enumitem}
\newlist{mycompactenum}{enumerate}{1}
\setlist[mycompactenum,1]{nosep,label=\arabic*.}
\usepackage{paralist}

\newcommand{\wfirst}{\textit{Roman}}

\title[Microlensing due to free-floating moon-planet systems]{Microlensing due to free-floating moon-planet systems}
\author[Sajadian et al.]{Sedighe Sajadian$~^{1}$\thanks{E-mail: s.sajadian@iut.ac.ir}, Parisa Sangtarash$~^{1}$\\
$^{1}$Department~of~Physics,~Isfahan~University~of~Technology,~Isfahan~84156-83111,~Iran}

\date{Accepted XXX. Received YYY; in original form ZZZ}
\pubyear{2023}

\begin{document}
\label{firstpage}
\pagerange{\pageref{firstpage}--\pageref{lastpage}}
\maketitle
\begin{abstract}	
Gravitational microlensing is a powerful method for detecting and characterizing free-floating planetary-mass objects (FFPs). FFPs could have exomoons rotating them. In this work, we study the probability of realizing these systems (i.e., free-floating moon-planet ones) through microlensing observations. These systems make mostly close caustic configurations with a considerable finite-source effect. We investigate finite-source microlensing light curves owing to free-floating moon-planet systems. We conclude that crossing planetary caustics causes an extensive extra peak at light curves' wing that only changes its width if the source star does not cross the central caustic. If the source trajectory is normal to the moon-planet axis, the moon-induced perturbation has a symmetric shape with respect to the magnification peak, and its light curve is similar to a single-lens one with a higher finite-source effect. We evaluate the \wfirst~efficiency for realizing moon-induced perturbations, which is $\left[0.002-0.094\right]\%$ by assuming a log-uniform distribution for moon-planet mass ratio in the range $\in\left[-9,~-2\right]$. The highest detection efficiency (i.e., $\simeq 0.094\%$) happens for Saturn-mass planets when moon-planet distance is $\sim 43 R_{\rm p}$, where $R_{\rm p}$ is the Saturn radius. Enhancing planetary mass extends the event's time scale and decreases the finite-source effect, but it reduces the projected moon-planet distance normalized to the Einstein radius $s(R_{\rm E})$ which in turn decreases the size of planetary caustics and takes them away from the host planet's position in close caustic configurations.     	
\end{abstract}


\section{Introduction}
One of the known populations in our galaxy is free-floating or wide-separation planetary-mass objects (FFPs). The first known method to discover FFPs is infrared (or near-infrared) imaging observations. This method is sensitive to young and massive planets rather in young, and nearby stellar clusters with star-forming regions \citep[see,  e.g., ][]{Zapatero2000,2012ApJScholz,2012ApJPena,2013ApJLiu,2022NatAsMiret}.

\noindent For instance, the first low-mass and unbound object, OTS~44, was discovered in 1998 through infrared measurements from a star-forming region in the constellation Chamaeleon \citep{Oasa_1999}. Its mass was estimated as $11.5~M_{\rm J}$ which is located on a boundary between brown dwarfs and planets. Here, $M_{\rm J}$ is the Jupiter mass. For this low-mass object, infrared photometry and spectroscopy observations with the Spitzer Space Telescope, the Herschel Space Observatory, and the Very Large Telescope have revealed the existence of an accretion and rotating disc which proposes it can develop into a planetary system \citep{1998SciTamura,2005ApJLuhman,2013AAJoergens}.

Another method to discern far and low-mass FFPs located in the Galactic disc or even its halo is gravitational microlensing \citep[see, e.g.,][]{2011NaturSumi,2017NaturMroz}. During a microlensing event, the light of a background star is temporarily magnified because of its alignment with a massive and compact (even dark) object, i.e., the so-called lens object \citep{Einstein1936,Liebes1964,Chang1979}. The magnification factor is similar in different filters, which causes easily discerning lensing from other variability sources \citep[e.g., ][]{Gaudi2012,2018Tsapras}. A FFP can be the lens object in a microlensing event. That event has a timescale as short as a few days, and may exhibit (sometimes extreme) finite-source effects \citep{1994WittMOA} because of large normalized source radii \citep{Mroz2018}.

\begin{table*}
\centering
\begin{tabular}{cccccccccc}\toprule[1.5pt]
$\rm{Targets}$& $\log_{10}\big[q\big]$& $d$ &$P$ &$\left<t_{\rm E}\right>$&$\log_{10}[\left<R_{\rm E}\right>]$ &$\log_{10}[\left<s\right>]\pm\log_{10}[\sigma_{s}]$&  $\left<\rho_{\star}\right>$ &  $\log_{10}[\left<\Delta_{\rm c}\right>]$ & $\log_{10}[\left<\Delta_{\rm p}\right>]$ \\
& & $(R_{\rm p})$ & $\rm{(days)}$ &$\rm{(days)}$& & $\rm{(R_{\rm E})}$ & & & \\
\toprule[1.5pt]
$\rm{Earth}$-$\rm{Moon}$ & $-1.910$   &  $60.34$  &  $27.29$  &  $0.104$  &  $-2.260$ & $-0.377\pm-0.590$   &  $0.555$  &  $--$  &  $-1.621$ \\
$\rm{Mars}$-$\rm{Phobos}$ & $-7.778$   &  $2.77$  &  $0.32$  &  $0.035$  &  $-2.749$ &$-1.501\pm-1.713$   &  $1.711$  & $-10.634$  &  $-7.927$\\
$\rm{Mars}$-$\rm{Deimos}$&$-8.636$   &  $6.92$  &  $1.27$  &  $0.034$  &  $-2.749$ & $-1.104\pm-1.316$   &  $1.678$  &  $-10.651$  &  $-7.164$ \\
$\rm{Jupiter}$-$\rm{Europa}$ & $-4.597$   &  $9.60$  &  $3.55$  &  $1.878$  &  $-1.013$ & $-1.383\pm-1.592$   &  $0.032$  &  $-7.210$  &  $-5.972$\\
$\rm{Jupiter}$-$\rm{Ganymede}$ &$-4.108$   &  $15.31$  &  $7.16$  &  $1.869$  &  $-1.013$ & $-1.179\pm-1.390$   &  $0.031$  &  $-6.289$  &  $-5.125$\\
$\rm{Jupiter}$-$\rm{Callisto}$&$-4.247$   &  $26.93$  &  $16.69$  &  $1.872$  &  $-1.012$ &$-0.935\pm-1.150$   &  $0.031$  &  $-4.063$  &  $-4.470$\\
$\rm{Jupiter}-\rm{Io}$  &  $-4.327$   &  $6.03$  &  $1.77$  &  $1.860$  &  $-1.011$ & $-1.586\pm-1.798$   &  $0.031$  &  $-7.356$  &  $-6.455$  \\
$\rm{Jupiter}-\rm{Amalthea}$  & $-8.960$   &  $2.59$  &  $0.50$  &  $1.873$  &  $-1.013$ &$-1.951\pm-2.162$   &  $0.032$  &  $-12.722$  &  $-9.864$\\
$\rm{Saturn}$-$\rm{Titan}$&$-3.626$   &  $20.98$  &  $15.95$  &  $1.021$  &  $-1.274$& $-0.861\pm-1.071$   &  $0.056$  &  $-3.806$  &  $-3.922$\\
$\rm{Saturn}$-$\rm{Rhea}$&$-5.392$   &  $9.05$  &  $4.52$  &  $1.016$  &  $-1.274$&$-1.228\pm-1.437$&  $0.058$  &  $-7.677$  &  $-5.904$\\
$\rm{Saturn}$-$\rm{Iapetus}$&$-5.498$ &$61.15$&$79.34$&$1.026$  &  $-1.274$ & $-0.396\pm-0.610$&  $0.056$  &  $-1.972$  &  $-3.475$ \\
$\rm{Saturn}$-$\rm{Dione}$&$-5.715$ &  $6.48$  &  $2.74$  &  $1.021$  &  $-1.274$ &$-1.370\pm-1.580$&$0.058$  &  $-8.302$  &  $-6.495$ \\
$\rm{Saturn}$-$\rm{Tethys}$&$-5.964$ &  $5.06$  &  $1.89$  &  $1.017$  &  $-1.274$&$-1.481\pm-1.693$&  $0.056$  &  $-8.779$  &  $-6.956$ \\
$\rm{Saturn}$-$\rm{Enceladus}$&$-6.721$& $4.09$  &  $1.37$  &  $1.011$&$-1.274$&$-1.570\pm-1.775$&  $0.058$  &  $-9.714$  &  $-7.592$\\
$\rm{Saturn}$-$\rm{Mimas}$ &$-7.181$ &  $3.19$  &  $0.94$  &  $1.019$  &  $-1.274$&$-1.681\pm-1.891$ &  $0.058$  &  $-10.399$  &  $-8.161$\\
$\rm{Uranus}$-$\rm{Titania}$&$-4.407$   &  $17.19$  &  $8.70$  &  $0.400$  &  $-1.683$& $-0.899\pm-1.112$   &  $0.148$  &  $-5.443$  &  $-4.439$\\
$\rm{Uranus}$-$\rm{Oberon}$&$-4.451$   &  $23.01$ & $13.47$ &$0.395$&$-1.683$ &$-0.773\pm-0.987$   &  $0.147$  &  $-0.805$  & $-4.081$\\
$\rm{Uranus}$-$\rm{Umbriel}$ & $-4.833$   &  $10.49$&$4.14$&$0.400$&$-1.683$&$-1.113\pm-1.321$&$0.149$  &  $-6.860$  &  $-5.282$\\
$\rm{Uranus}$-$\rm{Ariel}$& $-4.841$   &  $7.53$&$2.52$  &  $0.404$&$-1.682$&$-1.259\pm-1.471$   &  $0.146$  &  $-7.197$  &  $-5.731$\\
$\rm{Uranus}$-$\rm{Miranda}$&$-6.132$   &  $5.10$&$1.41$  &  $0.398$  &  $-1.682$&$-1.428\pm-1.637$   &  $0.147$  &  $-8.838$  &  $-6.879$\\
$\rm{Neptune}$-$\rm{Triton}$&$-3.680$   &  $14.41$&$5.88$  &  $0.436$  &  $-1.647$&$-1.025\pm-1.233$   &  $0.135$  &  $-5.471$  &  $-4.438$\\
$\rm{Neptune}$-$\rm{Proteus}$& $-6.367$   &  $4.78$  &  $1.12$  &  $0.432$  &  $-1.646$&$-1.506\pm-1.718$   &  $0.135$  &  $-9.233$  &  $-7.235$\\
$\rm{Neptune}$-$\rm{Nereid}$& $-6.519$  & $223.54$  &  $359.19$  &  $0.439$  &  $-1.646$&$0.164\pm-0.040$&$0.137$&$--$  &  $--$\\
$\rm{Pluto}$-$\rm{Charon}$&$-0.917$   &  $16.49$  &  $6.37$  &  $0.005$  &  $-3.556$&$-0.391\pm-0.692$   &  $10.177$  &  $--$  &  $--$\\
$\rm{Pluto}$-$\rm{Hydra}$& $-5.436$   &  $54.48$  &  $40.53$  &  $0.005$  &  $-3.580$&$0.150\pm-0.153$   &  $10.837$  &  $--$  &  $--$\\
\hline
\end{tabular}
\caption{The lensing parameters for the known moon-planet configurations in our solar system. Here, $q$ is the moon-planet mass ratio, $d$ is the moon-planet distance normalized to the planet radius ($R_{\rm p}$), $P$ is the moon orbital period, and $s$ is the projected moon-planet distance on the sky plane which is normalized to the Einstein radius ($R_{\rm E}$). $\sigma_{s}$ is the Standard Deviation of the $s$ distribution. Two last columns are rough estimations of the central and planetary caustic sizes in the logarithmic scale, as derived by $\Delta_{c}= q\left(s-1/s \right)^{-2}$, and $\Delta_{\rm p}=\sqrt{q}~s^{3}$.}\label{tab1}
\end{table*}

Three microlensing surveys, including the Optical Gravitational Lensing Experiment (OGLE, \citet{OGLE_IV}), The Microlensing Observations in Astrophysics (MOA, \citet{MOA_gourp}), Korea Microlensing Telescope Network (KMTNet, \citet{KMTNet2016}), have frequently reported (and report) detecting FFPs in our galaxy \citep[see, e.g., ][]{2011NaturSumi,2017NaturMroz,Mroz2018,2019AAMroz,2020MrozAJ,2020ApJMroz,2021Ryu_1,2021AJKim}. To derive a precise estimation of their abundance, \textit{The Nancy Grace Roman Space Telescope} (\wfirst)~ telescope is planned to monitor the Galactic bulge during six $62$-day observing seasons in its $5$-year mission. \citet{2020AJJohnson} reported that the \wfirst~ telescope would detect $\sim 250$ FFPs with masses down to that of Mars (including $\sim 60$ with masses $\lesssim m_{\oplus}$).

Generally, FFPs could be either ejected from their host planetary systems (even with their moons) owing to planet-planet scattering and dynamic instability \citep{LISSAUER1987249,LAUGHLIN2000614,2006Boss,Veras2012,2018ApJHong,2019MNRASRabago}, or formed through core accretion inside protoplanetary discs in a similar way to stars or brown dwarfs \citep{IdaLin2004,2009Mordasini,2012luhman}. In the first mechanism, Earth-mass planets are ejected more frequently than Jupiter-like ones \citep{2015Pfyffer,Coreaccretion}. 

\indent We note that FFPs could have exomoons orbiting them which can be produced by either planet-planet collision or co-accretion, and capture \citep{Debes_2007,2016Barr}. Exomoons orbiting FFPs even potentially maintain liquid water on their surface which increases the probability of life formation over them \citep{1987AdSpReynolds,2006ApJScharf,2017AADobos,liquidwater}. This property of free-floating moon-planet systems promotes the importance of detecting them. In the regard of discovering such systems \citet{2021ApJLimbach} investigated the probability of detecting exomoons transiting FFPs, and concluded that detection of Io and Titan-like exomoons is possible by \textit{James Webb Space Telescope} (JWST) telescope.

In this work, we study the possibility of discerning FFPs orbiting exo-moons through microlensing observations, especially with the \wfirst~telescope. In this regard, \citet{2002ApJHan} noticed the disruptive finite-source effect on realization of moon-induced perturbations in planetary lensing events. Also, \citet{2010Liebig} simulated lensing events from triple systems (including a moon, planet, and host star) and concluded discerning such systems would be possible through frequent and highly precise monitoring of the Galactic bulge. Detecting such triple systems will be improved with the \wfirst's~ highly precise monitoring of the Galactic bulge \citep[see, e.g., ][]{2022AABachelet}.

In Section \ref{solar} we first review the known moon-planet systems orbiting the Sun. We assume they are free and make microlensing events, and study the properties of their microlensing events. In Section \ref{caucross} we investigate properties of caustic-crossing features in short-duration microlensing events due to free-floating moon-planet systems. In Section \ref{roman}, we evaluate the \wfirst~efficiency for detecting these systems during its Galactic Bulge Time Domain Survey. Finally, in Section \ref{conclu} we summarize the results and conclusions.

\section{Known moon-planet systems}\label{solar}
There are several moon-planet systems in our solar system. These known lunar systems are a good sample of possible free-floating moon-planets in our galaxy. Hence, we calculate the properties of their microlensing events that they would make if they were free. We assume these lunar systems are in our galaxy and in lines of sight toward the Galactic bulge and cause short-duration lensing effects for collinear and background stars located in the Galactic bulge. In order to have a correct sense of their lensing parameters, we calculate their average values over big ensembles of their  microlensing events. To simulate microlensing events due to each known moon-planet system, masses of planet and moon and their distance are fixed to the real values. Other parameters are determined according to the related distributions. For instance, the source photometry properties are chosen from the Besan\c{c}on model \footnote{\url{https://model.obs-besancon.fr/}} \citep{robin2003,robin2012}. The lens impact parameter, and the angle of source  trajectory with respect to moon-planet axis are chosen uniformly. Other details can be found in previous papers \citep[see, e.g.,][]{sajadian2019,Moniez2017,2015MNRASsajadian}.

In Table \ref{tab1}, the average values of lensing parameters due to the known moon-planet systems (mentioned in the first column) are reported. In this table $q$ is the moon-planet mass ratio, $d$ is the moon-planet distance normalized to the planet radius ($R_{\rm p}$), $P$ is the moon orbital period, $\left<t_{\rm E}\right>$ is the average Einstein crossing time on different possible microlensing events, $\left<R_{\rm E}\right>$ is the average Einstein radius (the radius of images ring at the complete alignment), and $\left<s\right>$ and $\sigma_{s}$ are the average projected moon-planet distance on the sky plane and normalized to the Einstein radius and the Standard Deviation from the average value, respectively \footnote{We use the notation proposed by \citet{2011ApJSkowron}.}. $\left<\rho_{\star}\right>$ is the average projected source radius on the lens planet and normalized to the Einstein radius.
 
\begin{figure}
\includegraphics[width=0.49\textwidth]{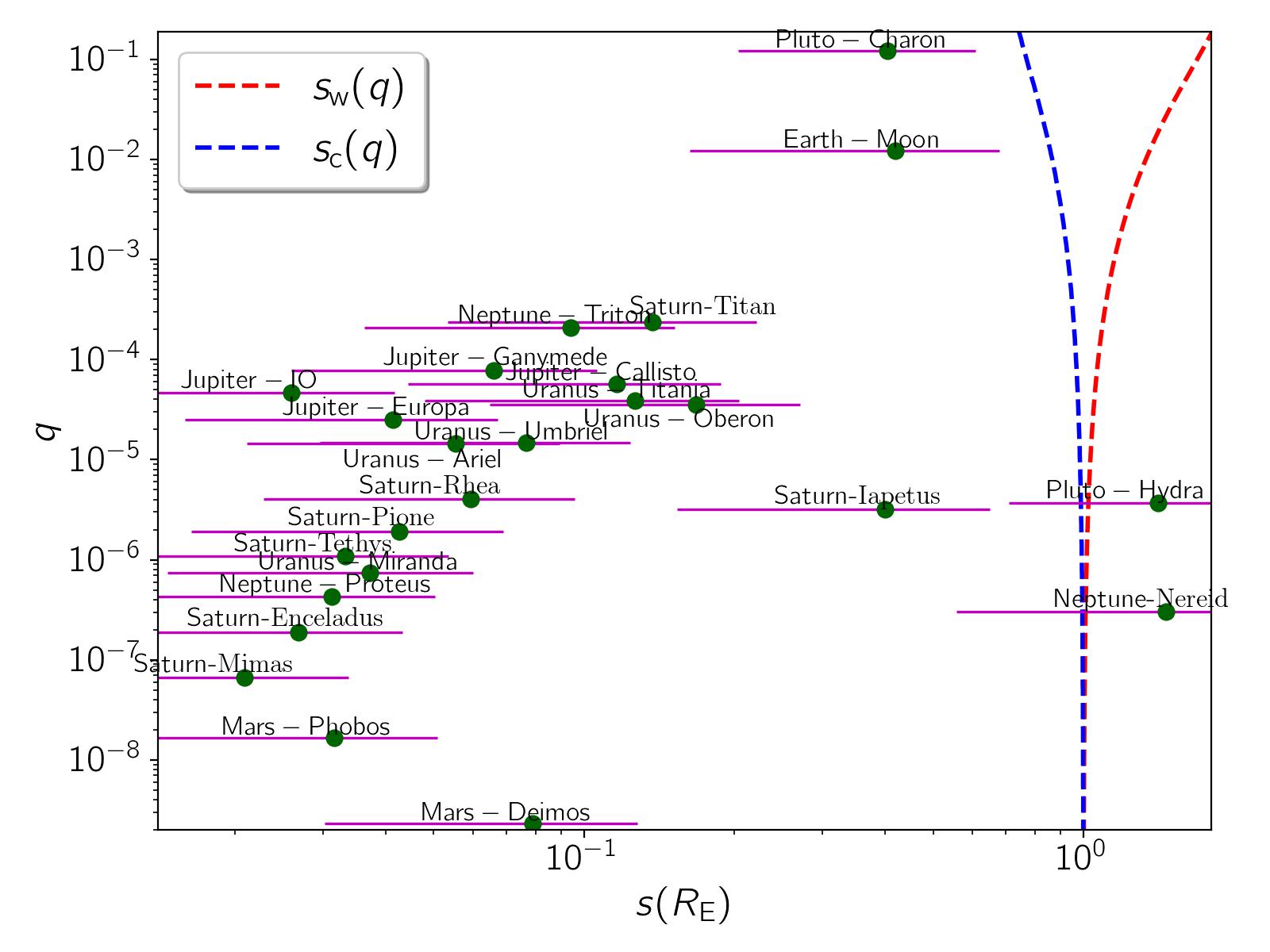}
\caption{The green points represent locations of the known moon-planet systems orbiting the Sun (as mentioned in Table \ref{tab1}) over 2D $s$-$q$ plane. The error bars in $s$ are the Standard Deviation from the average values in their simulated ensembles. $s_{\rm c}$ and $s_{\rm w}$ are the bifurcation values of $s$ between close-intermediate and intermediate-wide caustic topologies.} \label{fig1}
\end{figure}

\noindent Two last columns are the length scales of central and planetary caustics in the logarithmic scale which are roughly estimated by $\Delta_{\rm c}=q \left(s-1/s\right)^{-2}$, and $\Delta_{\rm p}= \sqrt{q}~s^{3}$ (respectively), for close caustic configurations and when $q\ll 1$ \citep{2000Bozza,2005An,Chung2005}. To determine their caustic configurations, in Figure \ref{fig1} we show positions of these moon-planet systems over 2D $q$-$s$ plane. The errors for normalized distance ($s$) are the Standard Deviation from the average values (seventh column of Table \ref{tab1}). Accordingly we list some key points in the following. 
\begin{itemize}
\item Most of these lunar systems make close caustic configurations with relatively small values of $q$.

\item Comparing the moon orbital periods and the average lensing timescales (mentioned in the fourth and fifth columns of Table \ref{tab1}), for most systems with large $q$ the lens orbital motion effect \citep[see e.g.,][]{2014MNRASsajadian} is ignorable in their binary lensing events, albeit they make close caustic configurations.

\item Similar to microlensing from FFPs, lensing due to lunar FFPs suffer from considerable finite-source effect. In all of these events the size of normalized source radius projected on the lens plane $(\rho_{\star})$ is at least 1-2 orders of magnitude larger than the size of caustic curves (comparing three last columns of Table \ref{tab1}). So, microlensing events due to free-floating moon-planet systems will not have obvious caustic-crossing features and the moon-induced perturbations due to passing over small caustic curves should be very small. 
\end{itemize}
In the next section, we characterize and evaluate the moon-induced perturbations in microlensing light curves. 

\begin{figure*}
	\subfigure[]{\includegraphics[angle=0,width=0.49\textwidth,clip=0]{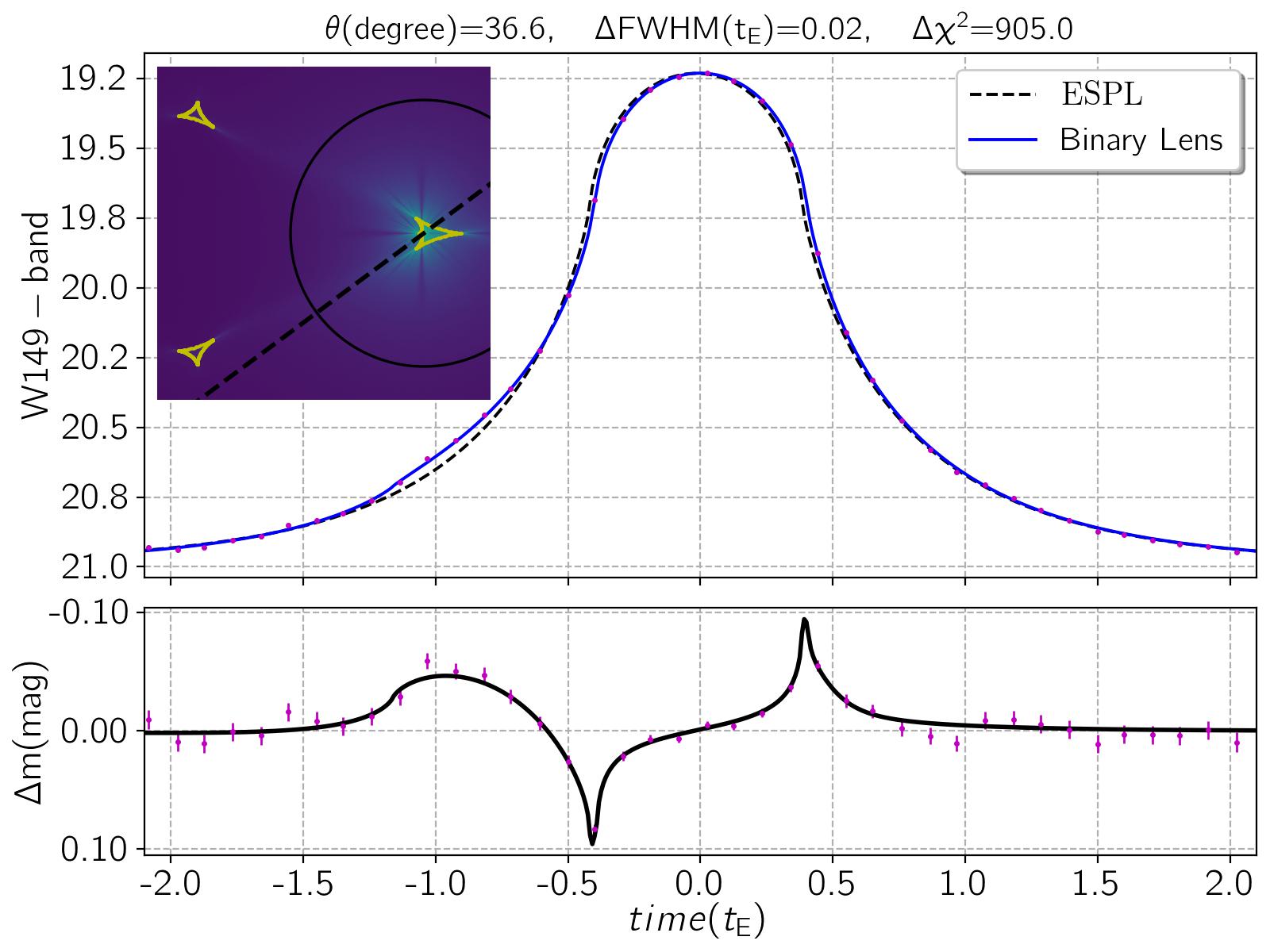}\label{la}}
	\subfigure[]{\includegraphics[angle=0,width=0.49\textwidth,clip=0]{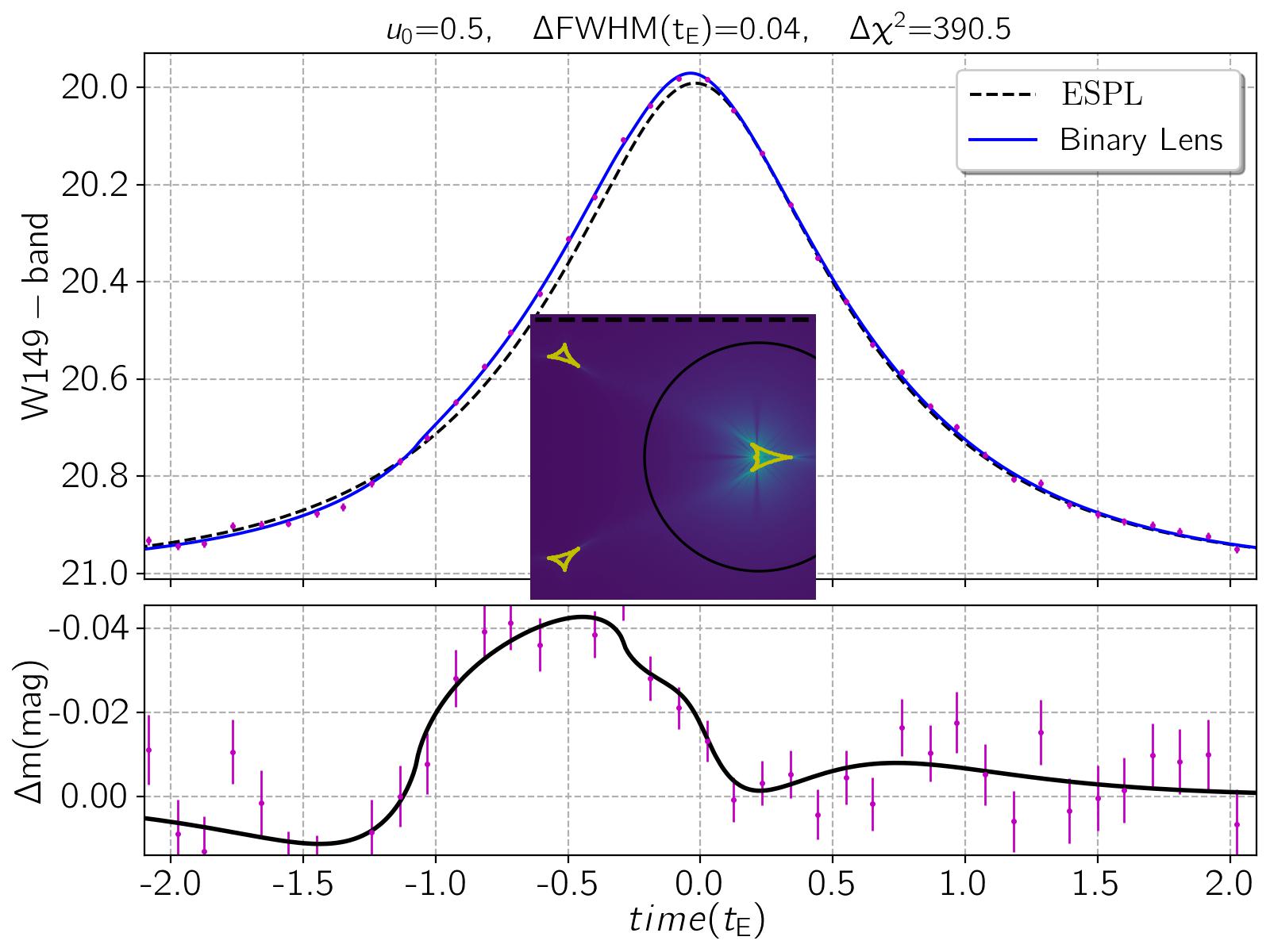}\label{lb}}
	\subfigure[]{\includegraphics[angle=0,width=0.49\textwidth,clip=0]{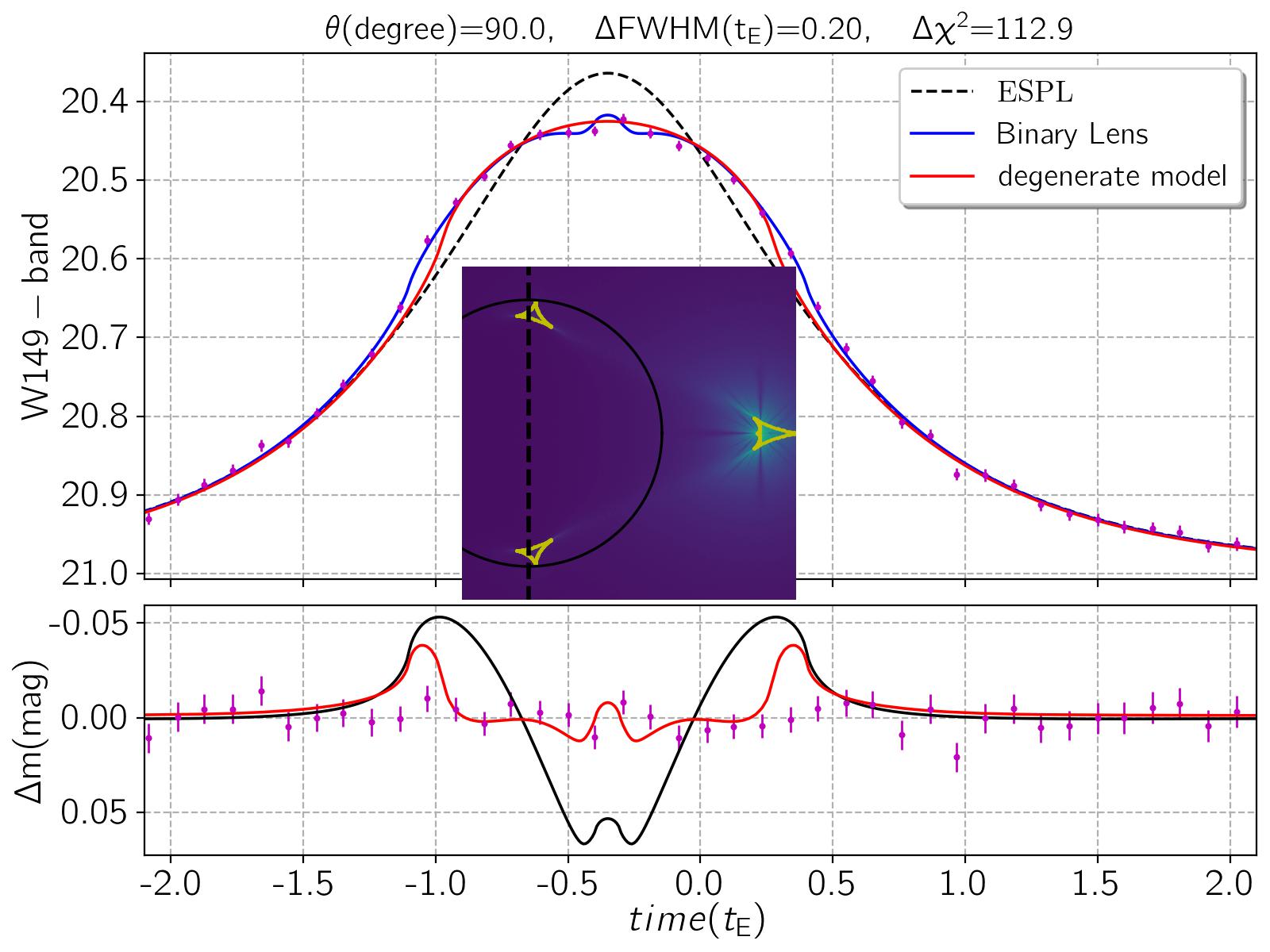}\label{lc}}
	\subfigure[]{\includegraphics[angle=0,width=0.49\textwidth,clip=0]{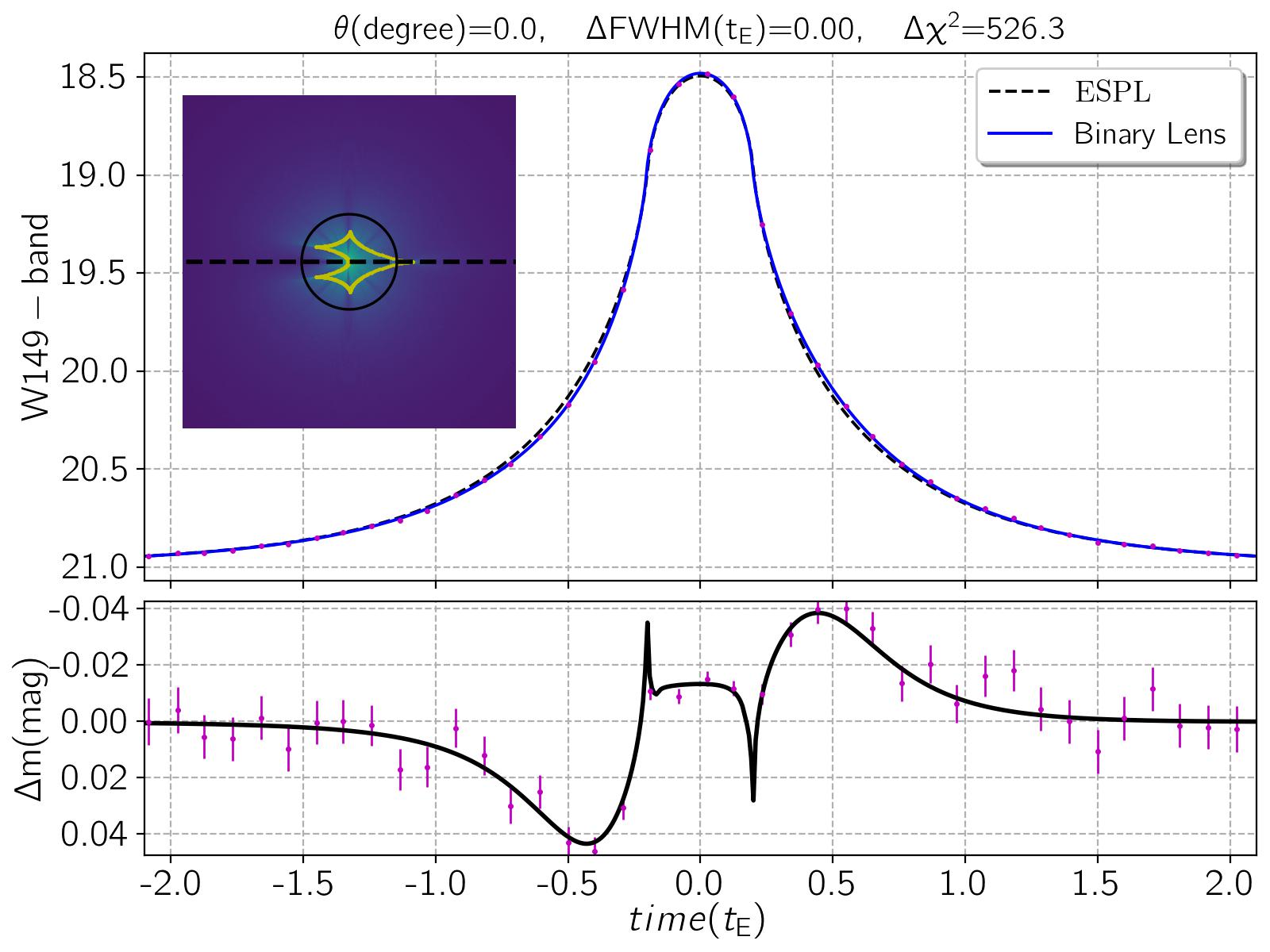}\label{ld}}	
	\caption{Four examples of microlensing light curves due to free-floating moon-planet systems with $q=0.03$, $s(R_{\rm E})=0.7$, $t_{\rm E}=0.1$ days, and $m_{\rm{base},~\rm{W149}}=21$ mag (solid blue curves). The corresponding single microlensing light curves are depicted by dashed black curves. The synthetic data points are simulated according to the \wfirst~observing strategy. The residual parts show moon-induced perturbations. The insets represent caustic maps, the source trajectory (dashed lines), and the normalized source sizes (solid black lines) projected on the lens plane.}\label{fig2}
\end{figure*}
\section{Caustic-crossing features}\label{caucross}
The \wfirst~telescope is planned to observe the Galactic bulge during six $62$-days seasons with the observing cadence $15.16~$min \footnote{\url{https://roman.gsfc.nasa.gov/galactic_bulge_time_domain_survey.html}}. During its mission, this telescope is predicted to detect $\sim 250$ FFPs with masses down to that of Mars (including $\sim 60$ with masses $\leq m_{\oplus}$) in short-duration and finite-source microlensing events \citep{2020AJJohnson}. In these events, the most significant second-order effect is finite-source, as a result of small Einstein radii. 

We note that short-duration microlensing events with extreme finite-source effect suffer from a continuous degeneracy which was first discerned by \citet{2020MrozAJ}, and then semi-analytically evaluated in \citet{2022Johnson}. Because of this degeneracy, there is a strong correlation between the lensing parameters $\rho_{\star}$, $t_{\rm E}$, $u_{0}$, and $f_{\rm b}$, so that they can not be realized uniquely. Here, $f_{\rm b}$ is the blending factor which is the ratio of the source flux to the total flux at the baseline, and $u_{0}$ is the lens impact parameter.

Additionally, if these FFPs host moon-like objects, another second-order perturbation disturbs their light curves. According to the previous section, free-floating moon-planet systems mostly make close caustic configurations that their caustic sizes are small compared to the normalized source radii. Hence, their light curves should be very similar to single and finite-source microlensing ones, but with small perturbations. Close caustic topologies include two planetary and one central caustic curves. Planetary caustics are larger and form at the opposite side of the secondary object's location and at the distance $\sim 1 /s - s$ from the location of primary lens along binary axis \citep[see, e.g., ][]{schneider1992,Gaudi2012}. 

In order to study the properties of such finite-source microlensing events due to free-floating moon-planet systems, we consider two moon-planet systems as a lens object and make several animations of their light curves due to planetary and central caustic crossings by varying (i) the lens impact parameter, (ii) the angle of source trajectory with respect to the binary axis, and (iii) the projected source size normalized to the Einstein radius. In these simulations, we make binary lensing light curves due to extended source stars using RT-model \footnote{\url{http://www.fisica.unisa.it/gravitationastrophysics/RTModel.htm}} developed by V.~Bozza \citep{Bozza2010,Skowron2012,Bozza2018}. All animations made in this work can be found at this address: \url{https://iutbox.iut.ac.ir/index.php/s/NYnnowMoLbFnZeD}.

\noindent In these animations, for each microlensing event two light curves are plotted in left panels: (a) real light curve of finite-source microlensing due to free-floating moon-planet systems (solid blue curve), and (b) single and finite-source microlensing light curve due to a lens with the total mass of planet and moon (dashed black curve). Their residual versus time normalized to the Einstein crossing time is represented in the bottom part. The caustic map, the projected source trajectory over it (black dashed line), and the source size projected on the lens plane and normalized to the Einstein radius $\rho_{\star}$ (solid black circle) are shown in the right panel. In these animations, in order to have a correct sense with respect to moon-induced perturbations and the difference between light curves, we assume that these events are detected by \wfirst~ in the W149 filter with the cadence $15.16$ min. The simulated data points are shown with magenta color. We generate the animations for two set of lensing parameters (I) $q=0.03$, $s(R_{\rm E})=0.7$, $t_{\rm E}=0.1$ days, $\rho_{\star}=0.4$, $m_{\rm{base},~\rm{W149}}=21$ mag (the source magnitude at the baseline ), and (II) $q=0.01$, $s(R_{\rm E})=0.42$, $t_{\rm E}=0.1$ days, $\rho_{\star}=0.56$, $m_{\rm{base},~\rm{W149}}=21$ (same as Moon-Earth system).

Four examples of these simulated light curves (which are shown in the animations) are represented in Figure \ref{fig2}. In the top of these light curves, the change in Full Width at Half Maximum (FWHM) of blue light curves with respect to black dashed ones normalized to the Einstein crossing time $\Delta \rm{FWHM}(t_{\rm E})$, and $\Delta \chi^{2}$ which is the difference between $\chi^{2}$ from fitting two (solid blue and dashed black) light curves to synthetic data points are reported. According to the animations, we list some key points in the following. 
\begin{itemize}
\item Generally moon-induced perturbations are very wide and their light curves are similar to single-lens ones without obvious perturbations.  

\item If the source star passes over planetary caustic in addition to crossing the central one, a very wide peak will appear in the wing of light curve (see Fig \ref{la}). Although the closest distance of the source trajectory to the central caustic determines the magnification peak, crossing the planetary caustic could change the light curve width. 

\item If the source star passes only over (or close to) planetary caustics, the peak location changes from the time of the closest approach to the central caustic. In that case, the moon-induced perturbations in light curves are small and barely realizable. One example light curve is shown in Figure \ref{lb}. However, these perturbations change the light curves width which leads to wrong estimations of $t_{\star}=t_{\rm E} \rho_{\star}$, i.e., the time of crossing the source radius.  

\item The least perturbations due to crossing planetary caustics happen when the source trajectory passes normal to the binary axis which leads to symmetric light curves with respect to the magnification peak. The width of these light curves is proportional to the normal distance between planetary caustics which is $\propto 2 q^{1/2} \sqrt{s^{-2}-1}$ \citep{2006Han}. This distance is significantly larger than the width of their corresponding single-lens light curve, i.e., $\propto t_{\star} \sqrt{1-u_{0}^{2}/\rho{\star}^{2}}$. Hence, these light curves are degenerate with single microlensing light curves by higher finite-source effects. For instance, in Figure \ref{lc}, one example of such microlensing events is shown. The degenerate model is plotted with red solid curve. The lensing parameters of this degenerate model is $\rho_{\star}=1.2$, $u_{0}=1.07$, and $t_{\rm E}=0.095$ days, whereas the lensing parameters of the real light curve (blue dashed curve) are $\rho_{\star}=0.4$, $u_{0}=0.7$, $t_{\rm E}=0.1$ days, $q=0.03$, and $s=0.7$. The difference between $\chi^{2}$ values for the real and the degenerate light curves is only $\Delta \chi^{2}=66$. Therefore, when the source trajectory passes normal to the binary axis, realizing the real model is almost impossible.  

 \item If free floating moon-planet system makes intermediate caustic, the highest moon-induced perturbation happens when the source trajectory is parallel with the binary axis. In that case, the light curve is not symmetric with respect to the magnification peak, although this asymmetry does not change the light curve width. One example is shown in Figure \ref{ld}.  
\end{itemize}

Accordingly, the moon-induced perturbations are mostly small and barely detectable. Nevertheless, the \wfirst~telescope with improved cadence $15.16$~min and high photometry accuracy may realize such small perturbations. We note that discerning a low-mass moon orbiting a free-floating planet far from us is only doable through gravitational microlensing observations. In next section, we study the \wfirst~efficiency for discerning moon-induced perturbations in finite-source microlensing light curves due to free-floating moon-planet systems.  
\begin{table*}
\centering
\begin{tabular}{cccccccccc}\toprule[1.5pt]
$M_{\rm p}$& $\left<t_{\rm E}\right>$& $\left<u_{0}\right>$ & $\log_{10}\left[\left<q\right>\right]$ & $\log_{10}\left[\left<s\right>\right]$ &  $\log_{10}\left[\left<\rho_{\star}\right>\right]$ & $\left<m_{\rm{base}}\right>$ & $\left<d\right>$ & $f_{\rm c}$:$f_{\rm i}$:$f_{\rm w}$ & $\epsilon(\%)$ \\
& $(\rm{days})$ & & & $(\rm{R_{\rm E}})$& & $(\rm{mag})$ & $(R_{\rm p})$ &  & \\
\toprule[1.5pt]
$M_{\oplus}$ & $0.42 \pm 0.45$ &  $0.21 \pm 0.25$  &  $-2.19 \pm -2.63$ &  $-0.18 \pm -0.87$ &  $-1.19\pm -1.43$ &  $19.6 \pm 1.4$ & $37.8 \pm 8.6$  & $90.9$:$0.0$:$9.1$ &  $0.00224$\\
$5~M_{\oplus}$ & $1.54 \pm 1.68$ &  $0.33 \pm 0.28$  &  $-2.39 \pm -2.53$ &  $-0.19 \pm -0.75$ &  $-1.14\pm -0.87$ &  $20.7 \pm 1.3$ & $36.8 \pm 6.0$  & $100.0$:$0.0$:$0.0$ &  $0.00243$\\
$10~M_{\oplus}$& $2.23 \pm 3.52$ &  $0.26 \pm 0.30$  &  $-2.18 \pm -2.60$ &  $-0.25 \pm -0.62$ &  $-1.10\pm -1.02$ &  $19.8 \pm 1.7$ & $42.7 \pm 8.7$  & $91.7$:$0.0$:$8.3$ &  $0.00405$\\
$50~M_{\oplus}$ & $2.05 \pm 2.69$ &  $0.30 \pm 0.28$  &  $-2.17 \pm -2.66$ &  $-0.25 \pm -0.78$ &  $-0.96\pm -0.98$ &  $19.9 \pm 1.9$ & $41.7 \pm 8.3$  & $92.9$:$0.0$:$7.1$ &  $0.04276$\\
 $100~M_{\oplus}$ & $2.14 \pm 2.91$ &  $0.41 \pm 0.31$  &  $-2.22 \pm -2.66$ &  $-0.21 \pm -0.63$ &  $-0.99\pm -0.92$ &  $19.6 \pm 1.8$ & $42.9 \pm 6.1$  & $95.7$:$4.3$:$0.0$ &  $0.08964$\\
 $150~M_{\oplus}$ & $2.37 \pm 4.48$ &  $0.34 \pm 0.29$  &  $-2.18 \pm -2.64$ &  $-0.21 \pm -0.62$ &  $-1.05\pm -1.11$ &  $19.8 \pm 1.6$ & $43.0 \pm 7.8$  & $90.3$:$2.8$:$6.9$ &  $0.09411$ \\
 $200~M_{\oplus}$ & $1.50 \pm 2.08$ &  $0.26 \pm 0.25$  &  $-2.16 \pm -2.62$ &  $-0.13 \pm -0.49$ &  $-1.04\pm -1.10$ &  $20.5 \pm 1.3$ & $42.1 \pm 8.8$  & $77.8$:$14.8$:$7.4$ &  $0.07919$\\
 $M_{\rm J}$ & $3.76 \pm 3.50$ &  $0.45 \pm 0.36$  &  $-2.17 \pm -2.69$ &  $-0.22 \pm -0.82$ &  $-1.32\pm -1.37$ &  $19.5 \pm 2.1$ & $48.0 \pm 3.2$  & $100.0$:$0.0$:$0.0$ &  $0.01841$\\
 $5~M_{\rm J}$ & $43.60 \pm 43.07$ &  $0.70 \pm 0.08$  &  $-2.07 \pm -3.19$ &  $-0.33 \pm -0.89$ &  $-0.87\pm -1.98$ &  $19.4 \pm 0.8$ & $53.7 \pm 0.3$  & $100.0$:$0.0$:$0.0$ &  $0.00254$\\
\hline
\end{tabular}
\caption{The average lensing parameters of microlensing events due to free-floating moon-planet systems with discernible moon signatures by \wfirst.~In the simulation, we consider discrete values for mass of planets as specified in each row. The last column is the \wfirst~efficiency for discerning moon-induced perturbations in these systems.}\label{tab2}
\end{table*}

\section{\wfirst~ Efficiency}\label{roman}
Up to now, no FFPs with rotating exomoons has been firmly confirmed through microlensing observations. However, the microlensing event MOA-2011-BLG-262Lb has a short time scale ($t_{\rm E}=3.8$ days) and a caustic-crossing feature. Although one possible solution for its lens is a free-floating moon-planet system, this event has another accidentally degenerate solution (a planetary system) which is favorable according to the Bayesian analysis \citep{Bennett_2014}. We expect that detecting microlensing events due to free-floating moon-planet systems considerably increases in the \wfirst~era. In this section, we focus on moon-planet systems and determine how the \wfirst~telescope is efficient to realize moon-induced perturbations.  

In this regard, we make big ensembles of microlensing events due to free-floating moon-planet systems. Then, synthetic data points taken by the \wfirst~ telescope are generated for each of them. Finally, we evaluate moon-induced perturbations in light curves and their detectability by \wfirst.~

We have little information about distribution functions of moon-planet systems, so we get the known moon-planet systems orbiting the Sun to help. In Table \ref{tab1}, parameters of 25 moon-planet configurations in our solar system were reported. In Appendix \ref{append1} we plot the fractional distributions of their $d/R_{\rm p}$, and $q$ in the logarithmic scale. In these plots, the average values and Standard Deviation from the averages are reported in plots with green color, and additionally depicted with dashed green (vertical) lines. These average values are close to the averages of log-uniform distributions (which are shown with black vertical lines in these figures). However, these two parameters have a correlation. This correlation is shown in the last panel of Figure \ref{dist}. Accordingly, there is a linear relation between $\log_{10}[d/R_{\rm p}]$ and $\log_{10}[q]$ as given by Equation \ref{eqcor}.
Hence, in simulations we first choose the moon-planet mass ratio uniformly in the logarithmic scale from the range $[-9, -2]$. Then, we determine the boundaries of all possible values for $\log_{10}[d/R_{\rm p}]$ (as depicted with magenta filled area in the last panel of Figure \ref{dist}, and given by Equation \ref{eqcor}). The final value for $\log_{10}[d/R_{\rm p}]$ is a random value between boundaries. 

To project the moon-planet distance on the sky plane, we consider a projection angle $\theta$. We choose $\cos \theta$ uniformly from the range $[0,~1]$. The projected moon-planet distance is $d~\sin \theta$.
 
We take other parameters including the photometry properties of source stars, extinction map, source distances, the lens distance, and their velocity components from the known distributions as explained in the previous papers \citep[see, e.g., ][]{sajadian2019,Moniez2017,2015MNRASsajadian}. To generate synthetic data points we determine the photometry error bars of data points according to the apparent magnitude of source stars in the W149 filter \citep{2020AJJohnson,2021sajadianSen}. The stellar absolute magnitude in W149 is roughly one-third of the summation of its absolute magnitudes in K, H, and J-bands \citep{2019MNRASbagheri,2020MNRASsajadian}. We assume that these events fully occur in one observing season. 

In previous section, we conclude the moon-induced perturbations should be small and very wide. Most of these perturbations make asymmetric features in light curves unless the source trajectory passes normal to the binary axis. Hence, to extract detectable moon-induced perturbations we apply two criteria which are (i) the difference between $\chi^{2}$ from fitting real models ($\chi^{2}_{\rm{real}}$), and single-lens and finite-source microlensing models (due to an object with the total mass of planet and moon ($\chi^{2}_{\rm{ESPL}}$) should be large, $\left|\Delta \chi^{2}\right|>800$, and (ii) two sides of light curves with respect to the magnification peak should be different (i.e., asymmetric features). We evaluate these asymmetric features using:
\begin{eqnarray}
\Delta \mathcal{A}= \sum_{i=1}^{N} \Big(\frac{A(t_{i})-A(t'_{i})}{\sigma_{i}}\Big)^{2}-  N
\label{asym}
\end{eqnarray}

\noindent where, $A$ is the magnification factor, $t_{i}$, and $t'_{i}$ are two mutual times with the same time interval with respect to the time of magnification peak, and $N$ is the number of data points taken in one side of light curves. The error in the magnification factor is specified as $\sigma_{i}= A(t_{i})\big[10^{-0.4\sigma_{\rm{m}}}-1\big]$, where $\sigma_{\rm{m}}$ is the \wfirst~ photometric precision \citep[see, Fig. (4) of ][]{2019ApJSPenny}.

In simulation, we noticed that $\Delta \mathcal{A}>1200$ is sufficient to exclude symmetric light curves. This second criterion is necessary to exclude light curves with symmetric moon-induced perturbations (e.g., Figure \ref{lc}). Such events are mostly interpreted as single-lens and finite-source microlensing events.  

We repeat simulations for different values of planets mass as $M_{\rm p} =M_{\oplus},~5M_{\oplus}, ~10M_{\oplus},~50M_{\oplus},~100M_{\oplus}, 150M_{\oplus},~200M_{\oplus},~M_{\rm J},~5M_{\rm J}$. In Table \ref{tab2}, results from these simulations are reported.     
\begin{figure*}
\centering
\includegraphics[width=0.49\textwidth]{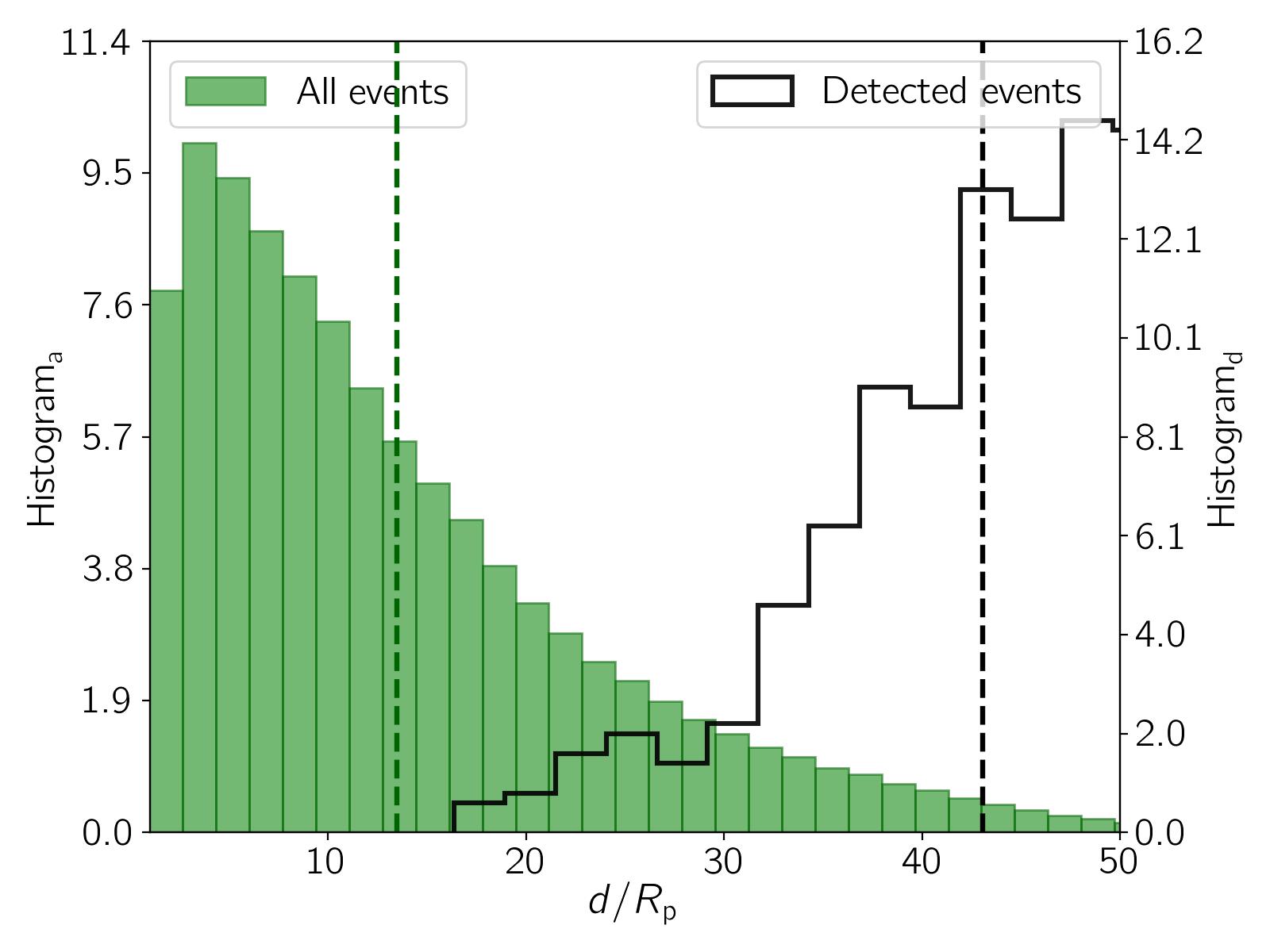}
\includegraphics[width=0.49\textwidth]{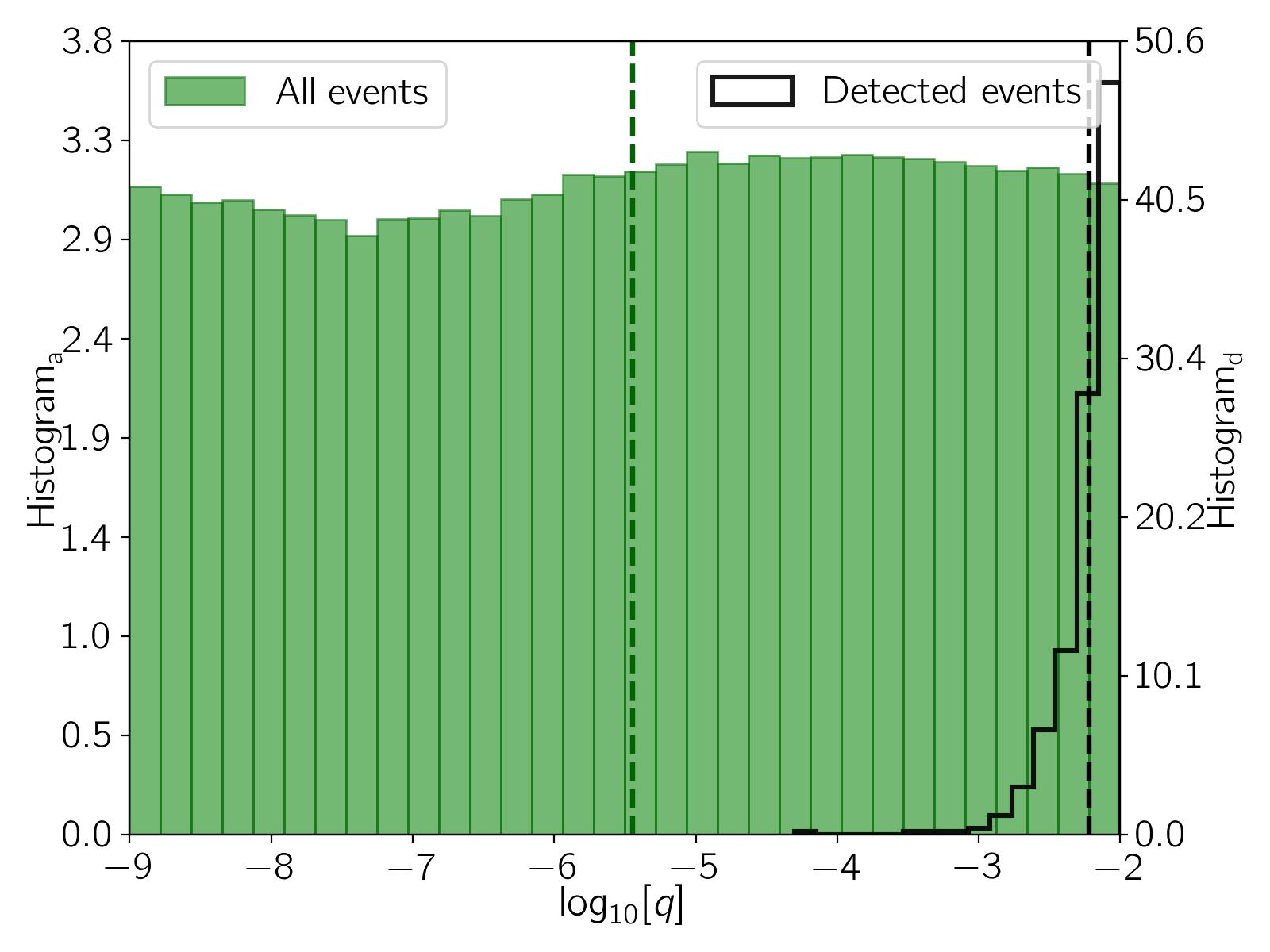}
\includegraphics[width=0.49\textwidth]{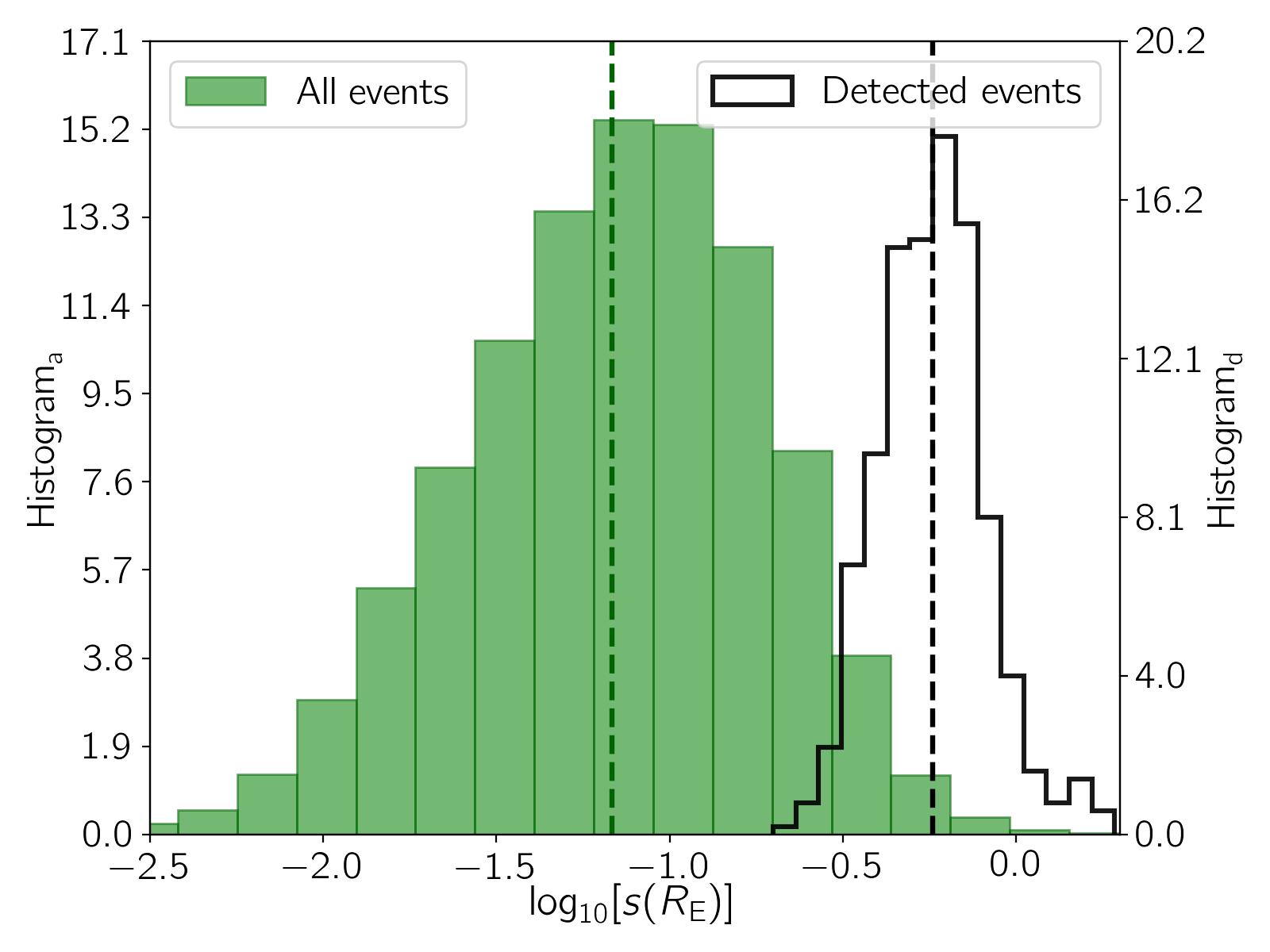}
\includegraphics[width=0.49\textwidth]{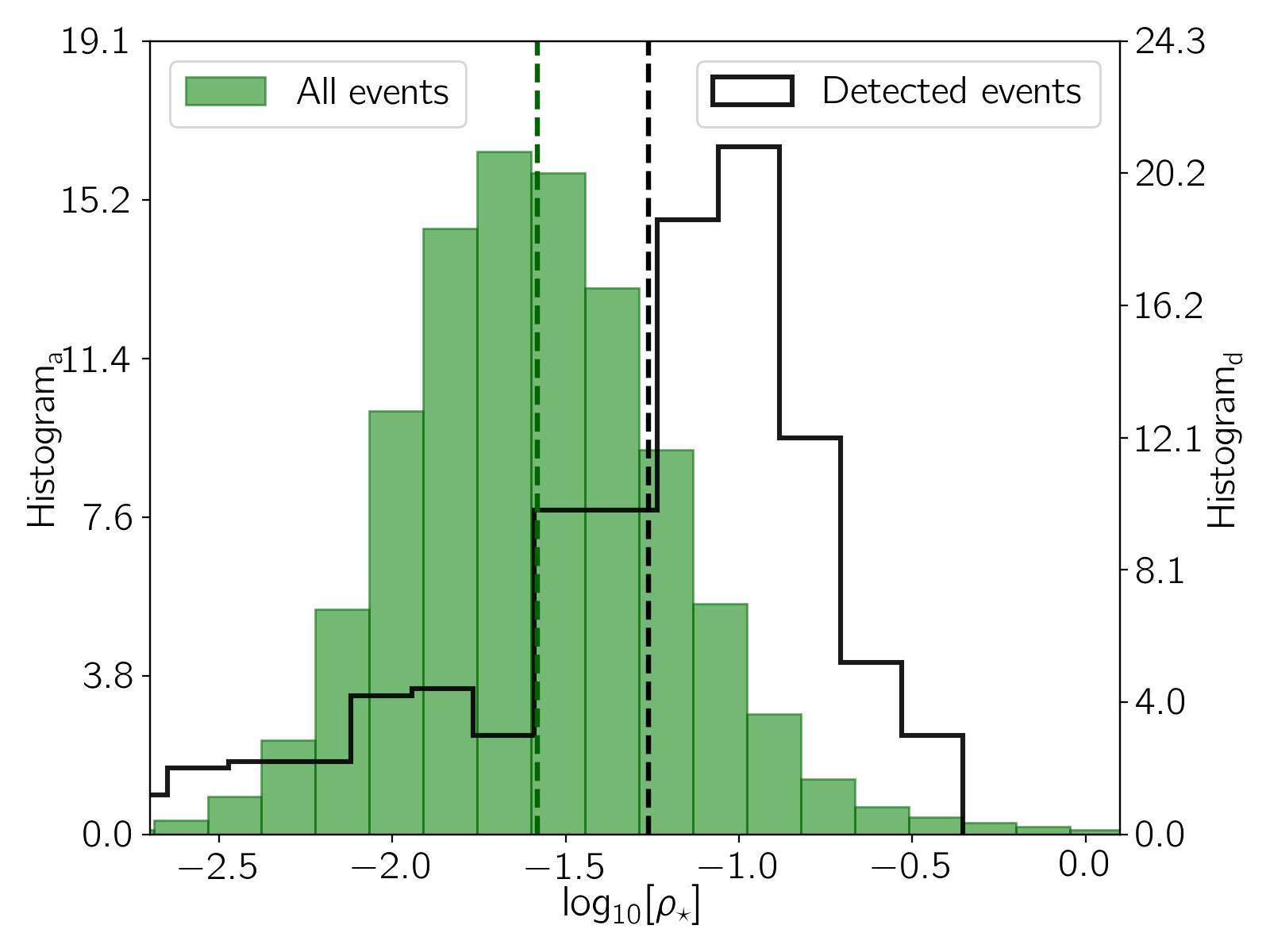}
\caption{The fractional histograms of four parameters $d/R_{\rm p}$,  $\log_{10}[q]$,  $\log_{10}[s(R_{\rm E})]$, and $\log_{10}[\rho_{\star}]$ for all simulated events (green filled histograms, specified on the left vertical axes) and detected events (black step ones, specified on the right vertical axes). These histograms are plotted for simulated events due to a planet with $M_{\rm p}=150M_{\oplus}$.}\label{histo}
\end{figure*}

According to Table, the \wfirst~telescope will detect moons around Earth-mass, super-Earth, Jupiter-mass, and super-Jupiter planets with efficiencies $\simeq 0.002\%,~0.004-0.094\%,~0.02\%,~0.002\%$, respectively. So the highest efficiency happens for moons orbiting Saturn-mass planets, while the moon-planet mass ratio is $\log_{10}\left[q\right]\simeq -2.18$, and their distance is $\simeq 43 R_{\rm p}$.\\ 

For detecting moon-induced perturbations in microlensing events due to free-floating moon-planet systems, four parameters have more impacts, which are (i) the moon-planet distance, (ii) the moon-planet mass ratio, (iii) the size and distance of planetary caustics from central caustics (which are functions of $s(R_{\rm E})$), and (iv) finite-source effect ($\rho_{\star}$). We show these points in Figure \ref{histo}. For events with $M_{\rm p}=150M_{\oplus}$, histograms of $d/R_{\rm p}$, $\log_{10}[q]$, $\log_{10}[s(R_{\rm E})]$, and $\log_{10}[\rho_{\star}]$ are plotted in each panel for all simulated events (green filled histograms) and detected events (black step ones). In these figures, green and black dashed lines determine the average values for green filled and black step histograms, respectively.

\noindent Increasing planets' mass reduces the finite-source effect and extends the event timescale, but it decreases the normalized moon-planet distance projected on the sky plane and as a result reduces the planetary caustic sizes. For that reason, the highest detection efficiency happens for planets with neither lowest nor highest masses. 
 
We note that the size of planetary caustics is proportional to $q^{1/2}~s^{3}$, their normal distance to the binary axis and their horizontal distance from the host location are $\propto 2 q^{1/2}\sqrt{s^{-2}-1}$, and $\propto s^{-1}-s$, respectively. We note that dividing $s$ and $q$ by two decreases planetary caustic sizes by $0.13$, and $0.7$, respectively. This shows a higher impact of $s$ than $q$ on planetary caustic topologies.

\section{Results and Conclusions}\label{conclu}
Free-floating planets (FFPs) could temporarily magnify the light of background and collinear source stars. The caused microlensing light curves have short timescales and usually are affected by finite-source effect. However, these FFPs can have moons orbiting them and even these exomoons potentially maintain liquid water. This ability increases the worth of detecting exomoons around FFPs. 

The \wfirst~telescope will do a time domain Galactic bulge survey during six 62-day seasons with a $15.16$~min cadence. The number of FFPs will be detected through the \wfirst~observations is $\sim 250$, as estimated by \citet{2022Johnson}. Some of these FFPs could have exomoons orbiting them developed by either planet-planet collision or co-accretion,
and capture \citep{Debes_2007,2016Barr}. In this work, we have studied how the \wfirst~telescope is efficient for realizing moon-induced perturbations in these microlensing light curves.  

We first studied the lensing parameters due to possible microlensing events made from 25 known moon-planet systems orbiting teh Sun. We concluded that most of them make close caustic configurations, with ignorable orbital motion effect of lenses, and considerable finite-source effect. So, these light curves should be similar to finite-source and single-lens events with very small moon-induced perturbations.  

By making several animations of light curves due to different configurations of free-floating moon-planet systems, we concluded that these perturbations are very wide, especially when source stars pass over (or close to) planetary caustic (see Figure \ref{la}). These wide moon-induced perturbations are barely detectable if the source star does not cross central caustic (low magnification). In that case, crossing only planetary caustics changes light curves's width, e.g., Figure \ref{lb}. 

If the source star passes normal to moon-planet axis, the caused light curves are symmetric and very similar to (degenerate with) single-lens ones with higher finite-source effect. One example is shown in Figure \ref{lc}. In this plot, the red solid curve is a degenerate and single-lens microlensing light curve that is well fitted to the synthetic data points. We note that the width of these light curves is proportional to the normal distance between planetary caustics. If free-floating moon-planet systems make intermediate caustic configurations, the largest asymmetric perturbations happen when the source star passes parallel with the binary axis.  

We also examined the \wfirst~efficiency for detecting moon-induced perturbations in light curves due to free-floating moon-planet systems by generating synthetic events and data points taken by \wfirst~in W149 filter. We consider discrete values for planet masses,  i.e., $M_{\oplus},~5M_{\oplus},~10M_{\oplus},~50M_{\oplus},~100M_{\oplus}, 150M_{\oplus},~200M_{\oplus},~M_{\rm J},~5M_{\rm J}$ and chose the moon-planet mass ratio from the range $\log_{10}\left[q\right]\in\left[-9,~-2\right]$ uniformly in the logarithmic scale. The moon-planet distances are chosen in the logarithmic scale according to $\log_{10}[q]$ (using Eq. \ref{eqcor}). We extracted detectable events by applying two criteria, (i) $\Delta \chi^{2}>800$ from fitting real and extended-source and single-lens models, and (ii) some asymmetry features should be in light curves, as defined in Equation \ref{asym}.    

We found that the detectability of moon-induced perturbations depends on several factors (i) the finite-source effect, (ii) the events' duration, (iii) the moon-planet mass ratio, and (iv) the size and distance of planetary caustics from central caustics. The first and second ones decrease by increasing the planet's mass. On the other hand, it reduces the projected moon-planet distance normalized to the Einstein radius. As a result, increasing the planet mass reduces the size of planetary caustics and enhances the distances of planetary caustics from central caustics. Moons orbiting Saturn-mass planets with $\log_{10}\left[q \right]\simeq -2.18$, and at the distance $\simeq 43 R_{\rm p}$ have the highest efficiency to be realized in the \wfirst~observations. These free-floating moon-planet systems make microlensing events with time scales of around $2.4$~days.  

\section*{Acknowledgements}
We specially thank the anonymous referee for her/his careful and useful comments.

\section*{DATA AVAILABILITY}
The animation made in this work can be found in \url{https://iutbox.iut.ac.ir/index.php/s/NYnnowMoLbFnZeD}. All simulations have been done for this paper are available at: \url{https://github.com/SSajadian54/Binary_FFPs_microlensing}.
\bibliographystyle{mnras}
\bibliography{paperref}
\appendix
\begin{figure}
	\centering
	\includegraphics[width=0.49\textwidth]{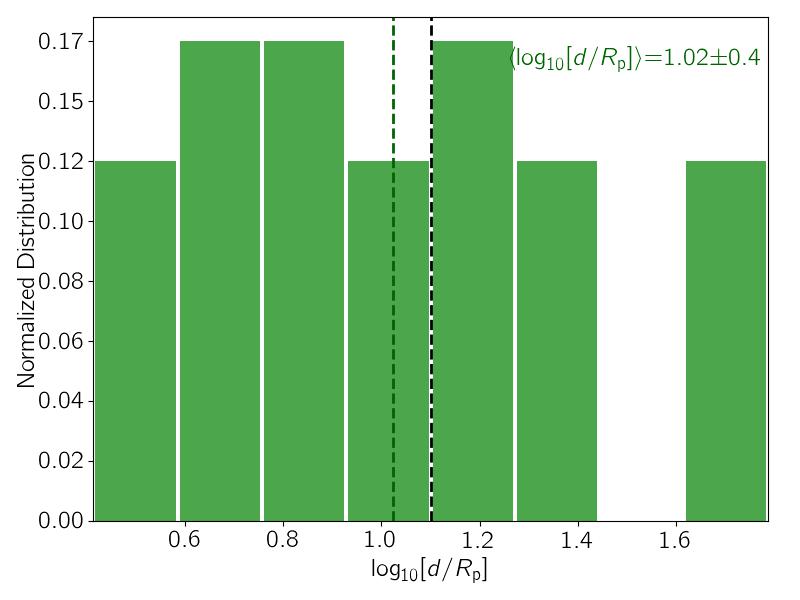}
	\includegraphics[width=0.49\textwidth]{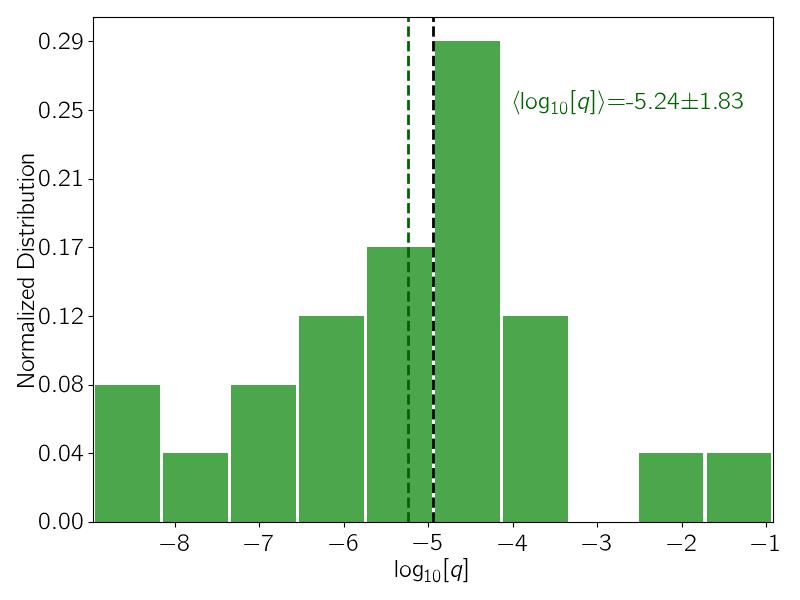}
	\includegraphics[width=0.49\textwidth]{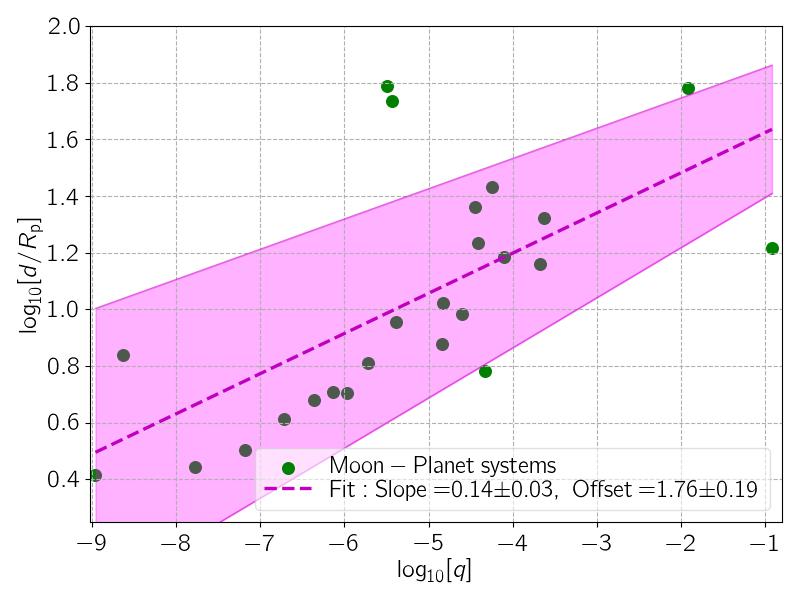}
	\caption{Two first panels are the normalized (fractional) distributions of $\log_{10}[d/R_{\rm p}]$, and $\log_{10}[q]$ of the known moon-planet binaries in our solar systems. The average values (depicted with dashed and green lines) and Standard Deviation from the averages are reported in plots with green color. The dashed and black lines are the average of log-uniform distributions of $\log_{10}[d/R_{\rm p}]$ in the range $[0.4,~1.8]$, and $\log_{10}[q]$ in the range $[-8.9,~-0.9]$. The characteristic of these binary systems are reported in Table \ref{tab1}. The last panel shows the dispersion of these systems over the 2D $\log_{10}[q]$-$\log_{10}[d/R_{\rm p}]$ plane.}\label{dist}
\end{figure}

\section{Distributions of the known moon-planet systems}
\label{append1}
In our solar systems there are many known moons orbiting planets. For instance, Earth has one moon, Mars has two moons (Phobos, Deimos), Jupiter and Saturn have 80 and 83 moons, respectively. A full list of these moons can be found in \url{https://solarsystem.nasa.gov/moons/overview/}. We list 25 of largest ones in Table \ref{tab1}. In Section \ref{roman}, we use their distributions to simulate microlensing events due to free-floating moon-planet systems. The normalized (fractional) distributions of the moon-planet distance normalized to the planet radius, and the moon-planet mass ratio in the logarithmic scale are shown in two first panels of Figure \ref{dist}, respectively. For each distribution, the average and Standard Deviation values are reported in plots with green color, and depicted with dashed and green lines. If we assume $\log_{10}[d(R_{\rm p})]$ and $\log_{10}[q]$ are uniformly distributed, their average and Standard Deviation values are $-4.94\pm2.23$, and $1.10\pm0.40$ (shown with dashed black lines). Comparing these average and Standard Deviation values, we can assume $d(R_{\rm p})$, and $q$ in the logarithmic scale are uniformly distributed. The Kolmogorov-Smirnov(K-S) test \citep{Handbook} for uniformity of these two distributions (of $\log_{10}[q]$ and $\log_{10}[d/R_{\rm p}]$) returns: K-S scores $0.25,~0.16$, and p-values $0.07,~0.48$, respectively. It means that K-S test confirms the uniformity of $\log_{10}[d/R_{\rm p}]$. But for distribution of $\log_{10}[q]$ its p-value is small (albeit it is higher than $0.01$) and K-S test can not firmly confirm its uniformity. However, the number of entrances is low (only 25). Additionally, we note that there is a correlation between these two parameters.

According to the last panel of Figure \ref{dist}, the dispersion of these moon-planet systems in 2D $\log_{10}[q]$-$\log_{10}[d(R_{\rm p})]$ plane is not uniform. On average, moons with larger mass ratios rotate in larger orbits and moons with mass ratios have smaller orbital radii. However, the scattering of data is relatively high and the number of entrances is low. Hence, this correlation is weak. We fit a linear relation between $\log_{10}[d(R_{\rm p})]$ and $\log_{10}[q]$ as:  
\begin{eqnarray}
\log_{10}[d/R_{\rm p}]= 0.14(\pm 0.03) \log_{10}\left[q\right] + 1.76(\pm 0.19).
\label{eqcor}
\end{eqnarray}	
Considering the errors of coefficients, for each given value of $\log_{10}\left[q\right]$ there is a range of possible values for $\log_{10}\left[d/R_{\rm p}\right]$, which is specified with a filled magenta region in the last panel of Figure \ref{dist}. In simulations of Section \ref{roman}, we use these distributions to simulate microlensing events due to free-floating moon-planet systems in our galaxy.
\end{document}